# Business information through Spain's Chambers of Commerce: meeting business needs

Antonio Muñoz-Cañavate

Pedro Hípola

Abstract:

From different public and private instances, mechanisms have been set in action that allow for companies to obtain information in order to make decisions with a stronger foundation. This article is focused on the description of an entire information system for the business world, developed in the realm of the Chambers of Commerce of Spain, which have given rise to the creation of an authentic network of inter-chamber information.

In Spain, the obligatory membership of businesses to the Chambers of Commerce in their geographic areas, and therefore the compulsory payment of member quotas, has traditionally generated some polemics, above all because many firms have not perceived a material usefulness of the services offered by these Chambers.

Notwithstanding, the 85 Chambers currently existing in Spain, as well as the organism that coordinates them —the Upper Council or *Consejo Superior de Cámaras de Comercio*— and the company created expressly to commercialize informational services online, Camerdata, have developed genuinely informative tools that cover a good part of the informational demands that a business might claim, described here.

**INTRODUCTION**

Over recent decades, Spain has witnessed a surge in very diverse tools that allow businesses to obtain information as an aid for enhanced decision-making. As in any other country, the organisms that produce information for the business world and that configure the informative structure underlying commerce are found in both the private and the public realm. Here, we shall take the private realm to mean that deriving from the private industry of information, whether it proceed from big businesses such as Kompass, Dun & Bradstreet, etc., or from the entire conglomerate of small national firms that offer information of very diverse sorts. Also included in the former group is the information proceeding from a great number of financial entities. In order to accommodate greater complexity, at times businesses that may be providers or distributors of products or services have managed to secure their influence by becoming the providers of information for a specific sector[1].

Yet the public bodies that supply information for the business world do not lag far behind, be they transnational networks such as that offered by the European Union, or any other organism that depends on the Public Administration of a given country.

In this latter sense, the regional structure of Spain offers a complex political-administrative panorama that has greatly diversified the informational tools made available to businesses, depending on the geographic region within Spain where they are located.

Aside from this, and as in other countries, Spain disposes of a network of Chambers of Commerce whose obligations for local business have permitted them to create a broad structure of very useful informational products.

**ORIGIN AND EVOLUTION OF THE CHAMBERS OF COMMERCE IN SPAIN**

The Chambers of Commerce, Industry and Navigation appeared in Spain at the end of the 19th century, thanks to a Royal Decree on April 8, 1886. Their birth was due to different causes —among them, the uncertainties among businessmen that sprouted with the appearance of the very first worker organization in the country. Furthermore, late 19th century Spain was characterized by an extension of the liberties of meeting and association, which favored the development of the Chambers of Commerce as a new framework for addressing business needs.

Spain opted to copy the French model of Chambers, to such an extent that the introduction of this norm in 1886 affirms:

> "France offers, with its Chambers of Commerce, an example that can, at least for now, be followed successfully. Created in the mid-17th century, they have been propagated by other nations, which already signal their advantages; and there is no reason why Spain should not also accept them as an anticipation of the epoch, just as long as importing the good that can be found in them, it takes care to mold them according to the general uses, customs and laws of the country".

The Royal Decree of 1886 recognized the Chambers as voluntary associations of shopkeepers, manufacturers and shippers. In 1901 they were acknowledged as public establishments, and in 1911 the model that continues to be in vigor regarding obligatory affiliation, with the payment of quotas, and under the auspices of the Public Administrations, was defined. Although the legal norm that has governed the Chambers of Commerce has varied over the decades, at present Spain's Chambers of Commerce are governed by Law 3/1993 (22 March, 1993), denominated *Ley Básica de las Cámaras Oficiales de Industria, Comercio y Navegación.*

This Law dating from 1993 states that the Chambers have the end purpose of representing, promoting and defending the general interests of commerce, industry and shipping, aside from lending services to business, without interfering with the liberty of syndication and of entrepreneurial association. In sum, the basic lines of activity of the Chambers are: actions to foment foreign commerce, formation, information and counselling.

The tasks involved in information and advisory are varied, and deserving mention among them are those dedicated to the elaboration of statistics regarding commerce, industry and shipping. They are also dedicated to the achievement of studies that allow us to be aware of the situation surrounding the different business sectors, and of course to carry out a public census of companies and their delegations. This has allowed the firms to generate very complete databases describing the situation of the economic and entrepreneurial setting[2].

The sections below will let us describe these informative tools.

## THE STRUCTURE OF THE CHAMBERS OF COMMERCE IN SPAIN, AND THE INFORMATION AND ADVISORY SERVICES

There exist in Spain 85 Chambers of Commerce, distributed over the whole of the national territory. Some of them cover designated areas such as the province (there are 52 provinces in Spain), whereas others cover a reduced territorial realm. Thus, they may have a local or provincial nature.

Legislation also regulates the existence of the Upper Council of the Official Chambers of Commerce, Industry and Navigation (*Consejo Superior de Cámaras Oficiales de Comercio, Industria y Navegación*), which coordinates the activity of the 85 Chambers, under the auspices, furthermore, of the Public Administrations.

The Upper Council represents, on a national and international level, the Chambers of Commerce of Spain, for example the representation in the International Chamber of Commerce or in the Euro-Chambers, aside from serving as spokesman and representative in the face of the Spanish Government. It also coordinates the activities of the Chambers, enhancing relations among them; and it reports on the position of the Chambers of Commerce before the government, from the information that, through its network, is transmitted by the business sector.

**The Study Service**

One of the work areas that is most important in this upper Council of Chambers is that of the Study Service, in charge of supplying economic information to businesses, in addition to elaborating a series of informational products that materialize in publications and databases, executed in an independent means or else in conjunction with the different study services of the Spanish Chambers. In some cases there is collaboration with other State organisms such as that dedicated to the promotion of exterior commerce (ICEX), and which interacts with services of different institutions, as is the case in other countries as well[3].

These informational products can be produced because the Law of the Chambers obliges them to elaborate statistics insofar as commerce, industry and shipping are concerned, in addition to having to carry out evaluation surveys and studies that make known the situation of the different sectors of the economy and of the business world.

The tables below give an overview of the studies and databases produced.

Table 1. Publications of the Study Service

|  |  |
| --- | --- |
| Economic analysis (studies that analyze economic situation and structural analyses). | Bulletin of the economic situation. Trimestral bulletins with the main outlooks regarding the international economy, that of the Euro zone, and that of Spain. |
|  | Annual economic report. It analyzes the behavior and the perspectives of international economics, the Spanish economy, and that of the regions of Spain. |
|  | Indicators of business confidence. Trimestral bulletins that are based on business surveys to expound opinions about the economic setting. |
|  | Perspectives for 2007 in Europe and Spain. A survey carried out by the Chambers of Commerce based on the opinion of tens of thousands of European entrepreneurs. |
|  | Perspectives 2007. Regarding the autonomous communities of Spain. |
|  | Perspectives 2007. Ibero-America. |
|  | Hispalink. Semestral reports. Spanish Regional Economic Outlook. It analyzes economic growth by regions and the main sectors for the year in course and the next two years. |
| Documents of foreign commerce | Reports on exporting and importing Spanish businesses. |
| Enterprise studies | Guides that aid in business management in a number of different aspects. |

Source: Consejo Superior de Cámaras de España.

Table 2. Databases of the Upper Council

|  |  |
|---|---|
| Directory of exporting and importing businesses | Directory of businesses with operations of foreign commerce, and the products they commercialize. |
| Database of foreign commerce | Statistics on international operations conducted by Spain (types of mechanisms, tons transported, number of operations to a destination). Data by province. |
| Regional database | 574 indicators of economic situation. The indicators show the evolution of economic activity of each region with respect to the Spanish national average. |
| National database: Info-Pais Cameral | Gives precise and simple information about the economic situation of the countries. A total of 162 countries and 51 variables sustaining the reports of each country. |
| Directory of sub-contracting Spanish businesses with export activity | Information about businesses that sub-contract in various sectors |
| National business census | Containing all the business firms associated with the Chambers |
| Sub-product markets | Gives the products that may be recycled for use by other businesses |

Source: Consejo Superior de Cámaras de España.

**The Business one-stop service**

The Upper Council of the Chambers also offers an advisory service known as "Ventanilla Única Empresarial", which might translate as "Business One-stop service". Its function is to support entrepreneurs in their attempts to create new enterprises.

This is an initiative of the State Administration of Spain, regional Administration, and local Administration, along with the Chambers of Commerce, offering services through the following mechanisms:

- Centers of contact (network available at http://www.ventanillaempresarial.org)
- The Business one-stop service (http://www.vue.es)
- Telephone for information
- E-mail.

**Legal advisory**

This service provides legal advice and counsel in specific matters regarding the environment. Within it we also find the Sub-product Market, which is a service that allows recyclable materials discarded by other businesses to be purchased for eventual use as sub-products.

### THE OFICIAL CHAMBERS OF COMMERCE, INDUSTRY AND NAVEGATION

As we mentioned above, there are 85 Official Chambers of Commerce, Industry and Navigation in Spain, their end purpose being the development of activities to support mercantile traffic and exterior commerce. The Chambers are connected with the commercial offices of Spain's embassies abroad, and dispose of different services for businesses, which collaborate with the Service of the Studies of the Upper Council of Chambers.

Traditionally, the Chambers have elaborated free information bulletins for their associates, although the rise of Internet has allowed them to develop broader information systems over the web.

Furthermore, the Chambers annually carry out censuses that are obligatory by Law (Ley 3/1993): on the one hand, the Census of Contributors, and on the other, the Statistical Census. Through these, commercial information can be facilitated to anyone who requests it.

In the Census of Contributors, the business openings that appear under the Economical Activity Tax (in Spanish, IAE) as to the 31 of December of each year are registered, whereas the Statistical Census contains breakdowns of this Tax under the

headings:

- Total activities classified by groups of activities.
- Total activities classified by geographic territory.
- Activities classified by groups of activity within each territory.
- Activities existing in each municipality.
- Number of activities in each IAE grouping.
- Activities classified by group within each municipality.
- Number of activities in each grouping, by municipality.

## CAMERDATA

The information system of the Chambers is complemented with a final informational tool: Camerdata, which stands as a corporation created by the Chambers themselves.

The origins of this incorporated entity can be traced to 1985 (just before the appearance and development of Internet in Spain), as a service that aspired to satisfy the incipient demand of information services for the business sector through online access.

Great expectations accompanied this project in its early years, due to the creation of one of the greatest databases for business in Spain (the Basic File), thanks to the compilation work of the Chambers. But the objectives were not totally covered**,** **s**o that the persons responsible for Camerdata were forced to work with a different outline, which meant reformulating objectives. These became:

- Introducing a new approach in the offer of information service for the Chambers.
- Renovation and updating of databases.
- The demand for new computer equipment.
- Greater international projection[4].

In this context, the following products appeared (available either by videotex, in Spain called Ibertex, or as an ascii service): International licitations (Linterna), economical situation and data from the Study Service of the Upper Council of Chambers of Commerce, Sub-product Market (offer and demand of industrial subproducts that allow the interchange of residue produced by one company that may be utilized by another), and Exportable Supply.

In 2007 the business information services of Camerdata were made up of the following structure of services, with their own databases plus other databases produced by other organisms, and distributed by Camerdata.1.

**1. Business File.** A database registering 2.5 million Spanish businesses (addresses of main offices and branches), over 2,000 economic activity codes and over 20 fields of information, in two blocks:

> 1.1. Basic data (name of company, address, postal code, municipality, province, main activity, other activities, legal form, headquarters or branch address, number of branches).

> 1.2. Additional data (telephone, fax, date of constitution, volume of business, number of employees, personal contacts and positions, and importing and exporting volumes).

**2. Commercial reports.** Linked to the Business File, they offer the following commercial information products:

> Basic file, company profile, stockholders and participations, main administrators, BORME acts (*Boletin Oficial del Registro Mercantil*), legal information, press articles, degree of solvency, basic report, commercial report, financial report, available balances and financial analysis.

**3. Sector reports**, with more than 200 sectors of activity (geographic and economic evolution of each sector).

**4. Information about all the businesses in a sector**. It holds data on Spanish businesses classified by sectors, regions and main rankings.

**5. National and European brands**. This database of national and European brands, commercial names, and signs of establishment, from 1990, includes the information registered in Spain's Office of Patents and Trademarks (*Oficina Española de Patentes y Marcas*).

**6. News items in national and economic press**. Daily coverage of news in the national and regional media.

**7. Grants and subventions.** A listing of national and European subventions. Also featured is a service of application and form processing, studies of possible subventions, and follow-up and negotiation of files.

**8. Public Works Competition**. Includes the public notice for licensing, openings and offers of public works, and concessions.

**9. The business in data**. It includes statistical reports elaborated in real time, from records found in the Spanish Business File.

**CONCLUSION**

We have described how, in Spain, the Chambers of Commerce have developed a genuine system of information that covers practically all the informational areas that may be of interest for a business person. These services, which may be free or fee-based, depending on the information one wishes to obtain, sustain a modern information system that can be accessed in full over the Internet.

[1] J. Tena Millán and A. Comai (2006). El Departamento de Inteligencia de Metalquimia, S. A. In *Inteligencia Competitiva y Vigilancia Tecnológica. Experiencias de Implantación en España y Latinoamérica*. Barcelona: EMECOM: 47-53.

[2] A. Valiente, E. García Duarte and A.Muñoz Cañavate (2005). La información para la empresa en las Cámaras de Comercio de Extremadura: un recorrido histórico. In A. Muñoz Cañavate (Coord.) *La información empresarial en Extremadura*. Badajoz: Diputación Provincial: 159-185.

[3] C. O'Hare (2004). Business Gateway. Business support in lowland Scotland. *Business Information Review* 21 (4): 245-251.

[4] A. Ulied (1993). La nueva Camerdata. *Information world en español* (*IWE*), 2 (14): p. 2-3.